\documentclass[A4paper,12pt]{article}

\usepackage{fancyhdr}
\usepackage{anysize}

\marginsize{1.5cm}{1.5cm}{1cm}{4cm}

\usepackage{amssymb}
\usepackage{graphicx}
\usepackage{amsmath}
\usepackage{amsfonts}
\usepackage{epsfig}
\usepackage{lineno}
\usepackage{color}
\newcommand{\sg}{\text{sg}}

\date{}
\begin{document}
\title{Quantization of BMS$_3$ orbits: a perturbative approach}

\author{Alan Garbarz$^{\dag,*}$, Mauricio Leston$^{\ddag,**}$}

\maketitle

\vspace{.2cm}

\begin{minipage}{.9\textwidth}\small \it \begin{center}
    $^\dag$ Instituto de F\'isica de Buenos Aires, CONICET $\&$ Departamento de F\'isica, Universidad de Buenos Aires,
Ciudad Universitaria, Buenos Aires, Argentina.
\\  \end{center}
\end{minipage}

\vspace{.2cm}

\begin{minipage}{.9\textwidth}\small \it \begin{center}
   $^\ddag$ Instituto de Astronom\'ia y F\'isica del Espacio, Pabell\'on IAFE-CONICET, Ciudad Universitaria, C.C. 67 Suc. 28, Buenos Aires
     \end{center}
\end{minipage}

\vspace{.5cm}

\begin{abstract}
We compute characters of the BMS group in three dimensions. The approach is the same as that performed by Witten  in the case of coadjoint orbits of the Virasoro group in the eighties, within the large central charge approximation. The procedure involves finding a Poisson bracket between classical variables and the corresponding commutator of observables in a Hilbert space, explaining why we call this a quantization. We provide first a pedagogical warm up by applying the method to both SL$(2,\mathbb{R})$ and Poincar\'{e}$_3$ groups. As for BMS$_3$, our results coincide with the characters of induced representations recently studied in the literature. Moreover, we relate the `coadjoint representations' with the induced representations.

\end{abstract}

\vspace{7cm}

\begin{flushleft}
\footnotesize
\parbox{\textwidth}{\mbox{}\hrulefill\\[-4pt]}
        {\scriptsize$^{*}$ E-mail: alan@df.uba.ar\\
\scriptsize$^{**}$ E-mail: mauricio@iafe.uba.ar}
\end{flushleft}

\newpage

\section{Introduction}

In the context of Conformal Field Theory in two dimensions and gravity theories in $2+1$ dimensions, the appearance of infinite-dimensional symmetries is inevitable and well understood. The task of investigating the infinite-dimensional groups of such symmetries  and their representations is as difficult as important. The main reason for the increasing difficulty is that procedures such as geometric quantization or the construction of induced representations stem on the use of certain invariant (or quasi-invariant) measures on some manifolds (coadjoint orbits), but these manifolds are infinite-dimensional  and such measures are not known (they may not exist in fact). Other issues also make things harder, as the fact that the standard methods for integrating a Lie algebra representation do not apply \cite{GoodmanWallach2}.

For the particular case of General Relativity in three dimensions one encounters the Virasoro group if a negative cosmological constant is considered, and the so-called BMS$_3$ group if no cosmological constant is present. These infinite-dimensional Lie groups appear as symmetries of the space of asymptotic solutions and it can be seen that such phase spaces are foliated by the coadjoint orbits of these groups \cite{GarbarzLeston,BarnichOblak}. The complete classification of Virasoro orbits was performed in \cite{Witten} and a more thorough study of the energy function as well as orbit representatives was done in \cite{Balog}. The same aspects where investigated in \cite{BarnichOblak, BarnichOblak1, BarnichOblak2} for BMS$_3$. 

A complete quantization of these spaces of solutions, roughly speaking, would start from considering the Poisson manifold that encompasses the union of coadjoint orbits as well as the classical observables on it, and then finding their corresponding quantum operators on a suitable Hilbert space where the classical symmetries are realized as unitary transformations. This is an open problem. What we can do instead is to concentrate on a sector of the classical theory, one coadjoint orbit. For example, in AdS$_3$ gravity, each BTZ black hole \cite{BTZ} lies in a unique Virasoro coadjoint orbit \cite{GarbarzLeston,BarnichOblak} of the type Diff$(S^1)/S^1$, and one can consider the observables on this symplectic manifold and attempt to give a unitary representation of them on some Hilbert space, thus quantizing this ``BTZ sector'' of AdS$_3$ gravity. It also has a mathematical interest on its own, and the complete classification of unitary positive-energy representations of Virasoro group has been recently given by Neeb and Salmasian in \cite{NeebSalmasian}. Their work can be regarded, among other things, as a (kind of) geometric quantization of the orbit Diff$(S^1)/S^1$, since their Hilbert space is given by certain holomorphic sections on a line bundle over the orbit. As far as we know, there is no analogue unitary representations for BMS$_3$ group\footnote{The BMS$_3$ induced representations studied in \cite{BarnichOblak1} rely on the unproven hypothesis that there exists a quasi-invariant measure on Virasoro orbits.}. 

Despite the difficulties mentioned at the beginning, it is possible to predict meaningful results without a full understanding of the quantum picture. Probably the most relevant one is the spectrum of the Hamiltonian, which can be obtained by computing the character of the time evolution operator, i.e. the partition function. For example, in \cite{Witten} this character of the Virasoro group is computed by means of a heuristic use of the Lefschetz formula \cite{AtiyahBott}. Remarkably, the same answer is given by a unitary Verma module representation of the Virasoro algebra\footnote{Although rarely mentioned, it is also a fact that using the Goodman-Wallach unitary representation of Virasoro group \cite{GoodmanWallach2} over the completion of a unitary Verma module (under its usual inner product) the character is still the same as that coming just from the algebra representation, since the Verma module is trivially dense on its completion.} and also by a perturbative analysis which turns out to give a system of free bosons \cite{Witten}.

As far as we know, there is no general argument explaining why the perturbative quantization (summarized below) gives the same characters as the ones obtained from non-perturbative methods. If it turns out to be the case that it suffices to study the symmetry group in a perturbative fashion in order to compute the characters, then it could also be the case that some other relevant aspects of the theories are fully accessible  already at the perturbative level. We are not going to explore this possibility here, but use it as a speculative additional reason to justify the perturbative approach we employ to study the quantization of certain orbits of BMS$_3$.

The perturbative method  for Virasoro orbits of \cite{Witten} (see also \cite{conical} for Diff$(S^1)/$PSL$^{(n)}(2,\mathbb{R})$ orbits), can be used in order to obtain a perturbative quantization of a particular orbit of a different group. We are going to exploit this in order to do such a thing for orbits of BMS$_3$. This procedure, in a few words, starts by putting some coordinates around a point in the orbit and writing the observables in a series expansion of these coordinates. Then, the algebra of observables is satisfied at leading order, and generically can be written as many (may be infinite) copies of Heisenberg algebra. Thus, a unitary representation can be obtained and in particular the characters can be computed. As a cautionary note, it should be said that characters associated to infinite-dimensional unitary representations (as the ones appearing in non-compact groups such as Poincar\'{e} group) are not well-defined as traces, and we will work following the lines of the results obtained in the late 60's for Poincar\'{e} group \cite{JoosSchrader}\footnote{See also \cite{FuchsRenouard} for a similar treatment of the characters.}. In that reference they regard the characters as distributions on the group algebra and analyse suitable spaces of test functions. We shall not go into this here, but it should be kept in mind in what sense we consider the characters of non-compact groups. 

The characters of BMS$_3$ have been studied recently, first in \cite{Oblak} and then in \cite{Barnichetal}. In the former the author considers the induced representations of BMS$_3$, labelled by mass $m$ and spin $j$, which assume the existence of a quasi-invariant measure on the Virasoro orbit Diff$(S^1)/S^1$, and then by means of Frobenius formula the author finds that 
\begin{equation}\label{Blazacharacter}
\chi_{j,m}(f,\alpha)=e^{ij\theta+i\beta m} \prod_{n\geq 1}\frac{1}{ |1-q^n|^2},\qquad q:=e^{i\theta},
\end{equation}      
where $f \in \text{Diff}(S^1)$ is conjugate to a rigid rotation by an angle $\theta$ and $\alpha \in \text{Vect}(S^1)$ has zero-mode equal to $\beta$. In complete analogy the character for representations obtained from the vacuum orbit Diff$(S^1)/\text{PSL}(2,\mathbb{R})$ is \cite{Oblak},
\begin{equation}\label{Blazacharacter2}
\chi_{j,m}(f,\alpha)=e^{ij\theta+i\beta m} \prod_{n\geq 2}\frac{1}{ |1-q^n|^2},\qquad q:=e^{i\theta},
\end{equation}  
The same character is obtained in the latter paper, where the authors use a functional approach to compute the euclidean partition function around Minkowski spacetime  for the case where $f$ is a rotation and $\alpha$ a time translation. 

In this note, as already anticipated, we will perform a perturbative quantization of massive and `vacuum' coadjoint orbits of BMS$_3$ group \footnote{The term `vacuum' comes from the fact that this particular orbit contains Minkowski spacetime in the context of flat three-dimensional gravity. Mathematically it possesses an enhacement of the little group, from $S^1$ to PSL$(2,\mathbb{R)}$.}. At the end, we will see that the relevant Hilbert space will be that of infinite two-dimensional non-relativistic free particles, i.e. an infinite tensor product of L$^2(\mathbb{R}^2)$ spaces. This Hilbert space will be identified with the one coming from induced representations and thus we will establish a concrete relation between the orbits (perturbative) quantization and induced representations  for BMS$_3$, in analogy with the case for finite-dimensional (nilpotent) groups. In particular, the characters we obtain coincide with (\ref{Blazacharacter}) and (\ref{Blazacharacter2}).

The presentation is organized as follows. In Section 2 we explain the procedure to find, in a perturbative manner, representations of the non-compact groups SL$(2,\mathbb{R})$ and $2+1$ Poincar\'{e} group departing from their coadjoint orbits. The advantage of using this perturbative procedure applied to these well-studied groups is that will ease the way towards the case of BMS$_3$ group. In Section 3 we turn into the BMS$_3$ group: we comment on the relevant features, mostly about its coadjoint orbits, and perform the perturbative quantization of massive and vacuum orbits. In Section 4 we discuss the results and comment on possible future lines of research.

\section{Warming up with SL$(2,\mathbb{R})$ and Poincar\'{e}}

In this section we will explain the procedure we will use later for BMS$_3$, in order to construct perturbative representations of the group and compute its characters. Instead of giving a general explanation, we will present here the approach for two pedagogical examples: SL$(2,\mathbb{R})$ and Poincar\'{e} in $2+1$ dimensions. The latter being the semi-direct product of the former with its algebra. In the next section, we will use the results obtained for the Poincar\'{e} group, so it will prove of much importance to our goals. We will not review basic facts about coadjoint orbits of Lie groups; the reader may want to consult \cite{GarbarzLeston,Witten} for short and simple summaries on the subject.

\subsection*{The perturbative quantization of SL$(2,\mathbb{R})$ orbits}
  
Let us consider first the group SL$(2,R)$, with its algebra $\textsf{sl}(2)$ generated by vectors $t_\mu$, $\mu=0,1,2$ such that
$$[t_\mu,t_\nu]=\epsilon_{\,\,\,\mu\nu}^{\lambda} t_\lambda,\qquad \epsilon_{012}=1.$$
The dual elements $t^{\mu *}$ can be defined by $\langle t^{\mu *},t_\nu\rangle=\delta^\mu_\nu$ and a generic coadjoint vector is of the form $q_\mu t^{\mu *}$. The coadjoint  orbits are hypersurfaces
$q_\mu q_\nu \eta^{\mu\nu}=-m^2$, with $\eta=\text{diag}(-1,1,1)$. We will focus in the one-sheeted hyperboloid $m>0$, with the point $q=m {t^0}^*=(m,0,0)$ being invariant under coadjoint transformations generated by $t_0$, i.e. rotations around the axes. 

Any coadjoint orbit is a symplectic manifold. We want now to describe the symplectic structure in terms of suitable coordiantes on the orbit. For this, take any adjoint vectors $u$ and $v$ in $\mathfrak{sl}(2)$ and write them as $u=x^\mu t_\mu$ and  $v=y^\mu t_\mu$. Thus, $\left\{x^\mu\right\}_{\mu=1,2}$ are  coordinates in a small neighborhood of the tangent space at a point in the orbit\footnote{Through the adjoint action, we identify adjoint vectors of the Lie algebra with vectors tangent to the orbit at some point.}. Then,  a  symplectic structure can be defined at $q$,
$$\omega_{(m,0,0)}(u,v):=-\langle (m,0,0),[u,v]\rangle=m(x^1 y^2-x^2 y^1).$$
This symplectic structure can be inverted in order to obtain the Poisson brackets in the coordinates
\begin{equation}\label{poissonsl2}
\left\{x^1,x^2\right\}=-m^{-1}.
\end{equation}
Note that this means that the coordinates are of order $1/\sqrt{m}$, and invites to think we are doing an expansion for $m>>1$. The Poisson structure close to the point $q$ is still given by (\ref{poissonsl2}) since any change $q+\delta q$ would contribute to higher orders in the expansion of the symplectic form. Then, we have a description of the symplectic structure around the point $q$ in the orbit.

We  would like now to be able to describe observables in the orbit, in particular in the region close to the point $q$. Even more, as usual, these observables should generate infinitesimal transformations associated to the algebra elements. For example, the observable $J_0=\Phi_{t_0}$ associated to a rotation in the hyperboloid generated by $t_0$ is given by
\begin{equation}
J_0(\text{Ad}^*_g(m,0,0)):=\langle  \text{Ad}^*_g(m,0,0), t_0\rangle=\langle  (m,0,0),\text{Ad}_{g^{-1}} t_0\rangle
\end{equation}
Similarly, $J_a(\cdot)=\langle \cdot, t_a\rangle$ for $a=1,2$. We can write $g=\exp(x^\mu t_\mu)$ and expand to second order the exponential\footnote{In this section we will disregard the fact that the exponential of the Lie algebra is not surjective. In the next section, when approaching the case of BMS$_3$, we will describe with suitable coordinates a patch close to the group identity and avoid the use of the exponential map.}, in order to get
\begin{equation}
J_0:=m+\frac{m}{2}\left(x_1^2+x_2^2\right).
\end{equation}
Then, given the fact that we can think of $x_1$ as the conjugate momenta of $x_2$, $J_0$ is the Hamiltonian of a harmonic oscillator. The two other observables read $J_1=-mx_2$ and $J_2=mx_1$, and together with $J_0$ they realize the $\textsf{sl}(2)$ algebra through the Poisson bracket at leading order in $1/m$.

Now that we have not only the symplectic structure in a neighborhood of $q$ but also the algebra of observables, we look for a quantization of this classical system. Taking the Poisson bracket (\ref{poissonsl2}), and defining $a:=\sqrt{\frac{m}{2}}(x_2+ix_1)$ we get upon quantization
$$[a,a^\dag]=1,\qquad J_0=m+a^\dag a.$$
Then, as anticipated, we have a quantum harmonic oscillator, which we can represent over the usual Fock space and the trace of $\exp(i\theta J_0)$ can be computed giving,
\begin{equation}
\text{Tr}(q^{J_0})=\frac{q^m}{1-q},\qquad q=e^{i\theta}.
\end{equation}
This is the character of the time evolution operator. 

\subsection*{The perturbative quantization of Poincar\'{e} orbits}

At this time we want to repeat what we just did for SL$(2,\mathbb{R})$ but for the Poincar\'{e} group in three dimensions. This is the semi-direct product 
$\text{SL}(2,\mathbb{R}) \ltimes_{Ad} \textsf{sl}(2,\mathbb{R})_{ab} $
and its algebra is $\textsf{sl}(2,\mathbb{R}) \ltimes_{ad} \textsf{sl}(2,\mathbb{R})_{ab}$, where the Lie product is 
$$[(X,\alpha),(Y,\beta)]=([X,Y],[X,\beta]-[Y,\alpha]) $$

Take again any two adjoint vectors $u$ and $v$ and write them as $u=(x^\mu t_\mu,\alpha^\mu t_\mu)$ and  $v=(y^\mu t_\mu,\beta^\mu t_\mu)$. Now the pair $(x^\mu,\alpha^\mu)$ gives the coordinates in a small neighborhood of a point in the orbit. On the other hand, the pairing of adjoint and coadjoint vectors is simply $$\langle (j,p),(X,\alpha)\rangle_{\text{Poincar\'{e}}}=\langle j,X\rangle_{\textsf{sl}(2)}+\langle p,\alpha\rangle_{\textsf{sl}(2)}.$$ 
The coadjoint orbits of a semi-direct product of Lie groups are revisited in \cite{BarnichOblak2}. We consider the orbit given by taking Ad$^*$ of the coadjoint vector $b_0=(jt^{0*},mt^{0*})$ with $j$ and $m$ real parameters (usually representing the spin and mass of the representation). Then, the symplectic structure is 
$$\omega_{b_0}(u,v):=-\langle b_0,[u,v]\rangle=j(x^1 y^2-x^2 y^1)-m(-x^1 \beta^2+x^2\beta^1+y^1\alpha^2-y^2\alpha^1)$$
This symplectic structure, thought as a $4\times4$ matrix, can be inverted in order to obtain the Poisson brackets in the coordinates. The non-zero brackets are
\begin{equation}\label{poissonpoincare}
\left\{x^1,\alpha^2\right\}=-m^{-1},\quad \left\{x^2,\alpha^1\right\}=m^{-1},\left\{\alpha^1,\alpha^2\right\}=j/m^2.
\end{equation}
Note that now $x^1$ and $x^2$ are not conjugate variables, and this can be understood from realizing that the orbits are cotangent manifolds (see \cite{BarnichOblak2} and references therein), where the base manifold is the hyperboloid with local coordinates $(x_1,x_2)$. On the other hand, the coordinates are all of order $1/\sqrt{m}\sim 1/\sqrt{j}$, so then again we take $m$ (and $j$) large.

In order to calculate some of the observables on the orbit, we need to remember what the group coadjoint  action looks like,
$$\text{Ad}^*_{f,\alpha}(jt^{0*},mt^{0*})=\left( \text{Ad}^*_f (jt^{0*})+\text{ad}^*_\alpha \text{Ad}^*_{f}(mt^{0*}), \text{Ad}^*_f (mt^{0*})\right),\qquad (f,\alpha)\in \text{SL}(2,\mathbb{R}) \ltimes_{Ad} sl(2,\mathbb{R})_{ab},$$
so the second argument is transformed as usual, while the first one gets modified by the presence of the vector $\alpha$. 
The observable associated to a translation generated by $(0,t_0)$ is given by
\begin{equation}
P_0(\text{Ad}^*_gb_0):=\langle\text{Ad}^*_gb_0,(0,t_0)\rangle=m+\frac{m}{2}\left(x_1^2+x_2^2\right).
\end{equation}
Where again we wrote $g=\exp(x^\mu t_\mu)$, and expanded to second order the exponential. 
In the same manner, the observable associated to $(t_0,0)$ is given by
\begin{equation}
J_0=j+\frac{j}{2}\left(x_1^2+x_2^2\right)+m\left(\alpha_1 x_1+\alpha_2 x_2\right)
\end{equation}

As already mentioned, the orbit is diffeomorphic to a cotangent bundle with local coordinates $(x_1,x_2)$. We now look for the conjugate coordinates. Define, $$p_1:=-\frac{j}{2} x_2 - m \alpha_2,\qquad p_2:=\frac{j}{2} x_1+m\alpha_1.$$ Then, we have a 2D particle, as should be the case for the orbits of three-dimensional Poincar\'{e} group: $$\left\{x_i,p_k\right\}=\delta_{ik}\quad i,k=1,2.$$

The complete set of observables $\Phi_X$ associated to an algebra element $X$ read:
\begin{eqnarray}
J_0&=&j+p_2 x_1-p_1x_2\\
J_1&=&p_1-\frac{j}{2}x_2\\
J_2&=&p_2+\frac{j}{2}x_1\\
P_0&=&m+\frac{m}{2}\left(x_1^2+x_2^2\right)\\
P_1&=&-mx_2\\
P_2&=&mx_1.
\end{eqnarray}
They satisfy $\left\{\Phi_X,\Phi_Y\right\}=\Phi_{[X,Y]}$ to first order in $m$ and $j$, so at this level of approximation the observables realize the algebra of Poincar\'{e}. Notice that $P_0$ is the typical free-particle Hamiltonian if the $x$-coordinates are thought as velocities while the $p$-coordinates as position coordinates. This also means that $J_0$ is the intrinsic spin plus an orbital angular momentum. 

We now turn to the task of quantizing this phase space and its observables, i.e. finding unitary representations for the transformations generated by such observables. The theory of unitary representations of the Poincar\'{e} group is well known \cite{Wigner}. In particular the characters have been studied in detail in \cite{JoosSchrader,FuchsRenouard}. We will show how to compute the characters of translations and rotations since this will smooth the arena for the BMS case. Choosing the typical Hilbert space $L^2(\mathbb{R}^2)$ and the representation given by $$x_k \psi=i\partial_{r_k}\psi(\vec{r}) \quad\text{and}\quad p_k \psi=r_k\psi,\qquad \psi \in  L^2(\mathbb{R}^2),$$ for $k=1,2$, we have for example
\begin{equation}\label{rotationandtranslation}
J_0=j+i r_2 \partial_1 - i r_1 \partial_2,\qquad P_0=m-\frac{m}{2}\nabla^2.
\end{equation}
In order to perform the computation of the character, we choose a basis of two-dimensional plane waves $\left\{\psi_{\vec{k}}\right\}_{\vec{k}\in\mathbb{R}^2}$. As is customary, we take the distribution-valued inner product $\left(\psi_{\vec{k}},\psi_{\vec{k'}}\right)=\delta(\vec{k}-\vec{k'})$. They are eigenfunctions of $P_0$ with eigenstate given by $m+\frac{m}{2}k^2$, while a rotation generated by $J_0$ acts on them as
\begin{equation}
e^{i\theta J_0}\psi_{\vec{k}}=\psi_{U_\theta \vec{k}},\qquad U_\theta=\left(
\begin{matrix}
\cos\theta & \sin\theta\\
-\sin\theta & \cos\theta
\end{matrix}
\right) \in \text{SO}(2). 
\end{equation}
In other words, a rotation generated by $J_0$ maps a plane wave to another plane wave with the wave vector rotated. Now we can compute the character associated to a $(j,m)$-representation:
\begin{eqnarray}\nonumber
\chi_{(j,m)}\left(e^{i\theta J_0+i\beta P_0}\right)&=&\int_{\mathbb{R}^2} d^2k \left(\psi_{\vec{k}},\left(e^{i\theta J_0+i\beta P_0}\right)\psi_{\vec{k}}\right)\\
&=&e^{i\theta j+i\beta m}\int_{\mathbb{R}^2} d^2k \left(\psi_{\vec{k}},\psi_{U_\theta\vec{k}}\right)e^{i\beta\frac{m}{2} k^2}\\
&=&e^{i\theta j+i\beta m}\int_{\mathbb{R}^2} d^2k\,\, \delta\left(\vec{k}-U_\theta\vec{k}\right)e^{i\beta\frac{m}{2} k^2}.
\end{eqnarray}
The exponential of $k^2$ does not matter since the delta distribution forces the only contribution to be the rotation-invariant vector $\vec{k}=0$. Moreover, the delta distribution contributes with a Jacobian $|1-e^{i\theta}|^{-2}$,
\begin{equation}\label{poincarecharacter}
\chi_{(j,m)}\left(e^{i\theta J_0+i\beta P_0}\right)=e^{i\theta j+i\beta m}\frac{1}{|1-q|^2},\qquad q=e^{i\theta}.
\end{equation}  
Notice that $P_0$, in our $m>>1$ aproximation, has the non-relativistic aspect of the energy. Going one order beyond in $1/\sqrt{m}$ should make relativistic corrections appear, but the `localization phenomena' in the character imposing $\vec{k}=0$ suggests that such relativistic corrections are not relevant. This deserves further investigation \cite{FarinatiGarbarzLeston}. Actually this fact resembles what happens for the characters of Virasoro coadjoint representations, where the parameter of the perturbation is the central charge $c>>1$ and, as it is shown in \cite{Witten}, it is likely the case that the functional form of the character of $q^{L_0}$ is  insensitive to the values of $c$.


\section{Perturbative quantization of BMS orbits}

In this section we want to apply the same procedure as before, i.e. for the Poincar\'{e} group, to the case of the BMS$_3$ group. This group is defined as the semidirect product of Virasoro group and its algebra (thought as an abelian group given the vector space structure). Note that there are two central extensions, one for each factor of the semidirect product. The \textsf{bms}$_3$ Lie algebra is the semidirect sum of two Virasoro algebras (see \cite{BarnichOblak1} and references therein). The coadjoint orbits of BMS$_3$ were discussed in \cite{BarnichOblak2}. We are interested first in the quantization of the massive particle orbits, i.e. the orbits where a coadjoint vector $b=(j,0;p,ic_2)$ is found. The orbits of this sort are labeled by the real parameters $j,p$ and $c_2$, which we take to be positive. 

\subsection{The symplectic structure of BMS orbits}
Let us first compute the symplectic structure. The adjoint vectors are generically of the form

$$u:=(X_1,-i\textsf{a}_1;\alpha_1,-i\textsf{b}_1) \in \textsf{bms}_3,\qquad v:=(X_2,-i\textsf{a}_2;\alpha_2,-i\textsf{b}_2) \in \textsf{bms}_3$$
As in the previous section, we define the symplectic form to be $\omega_b(u,v):=-\langle b,[u,v]\rangle$. Then in this case, 
\begin{equation}
\omega_b(u,v)=-\frac{ij}{\pi} \sum_{n\neq 0 }nf_n^1 f_{-n}^2-\frac{i}{\pi} \sum_{n\neq 0} 
n\left(p+\frac{c_2 n^2}{48\pi}\right)(f^1_n \alpha_{-n}^2+\alpha_n^1 f_{-n}^2)+\mathcal{O}(3)
\end{equation}
where $$X_i=\frac{1}{2\pi}\sum_{n\neq 0}f_n^i e^{-in\theta},\qquad \alpha_i=\frac{1}{2\pi}\sum_{n\neq 0}\alpha_n^i e^{-in\theta},$$
and we are considering only the leading terms in the $f_n$ and $\alpha_n$ coordinates, since $u$ and $v$ should be thought of as infinitesimal generators. Since we are interested in the tangent space to the orbit, we need the adjoint vectors in the quotient $\textsf{bms}_3/\textsf{u}(1)\oplus\mathbb{R}$, so we removed the zero modes from $u$ and $v$.  
It will prove useful to cast the pair of $f$- and $\alpha$-coordinates in a single set of coordinates,
\begin{equation}
T_{2n-\sg(n)}:=\alpha_n,\qquad T_{2n}:=f_n,\qquad T=(\ldots,\alpha_{-1},f_{-1},f_1,\alpha_1,\ldots) 
\end{equation}
where $\sg$ means the sign function. We have now a vector $T$ with all the coordinates. The symplectic form thus becomes,
\begin{equation}\label{symplecticform}
\omega(u,v)=-\frac{i}{\pi}\sum_{n\neq 0} n\left[j T^1_{2n}T^2_{-2n} + a_n\left(T^1_{2n} T^2_{-2n+\sg(n)}+T^2_{-2n} T^1_{2n-\sg(n)}\right)\right],
\end{equation} 
with,
$$a_n:=p+\frac{c_2 n^2}{48\pi},\qquad p\neq -\frac{c_2 n^2}{48\pi}. $$
The symplectic form (\ref{symplecticform}) has been computed at the point $b=(j,0;p,ic_2)\in \textsf{bms}_3$, but it actually retains its form in a neighborhood of $b$, since any $b'$ in this neighborhood will contribute to sub-leading order in the coordinates.
We can see from (\ref{symplecticform}) that the subindeces of $T^1$ and $T^2$ always sum $0,-1,1$, so arranged in a matrix they will form $2\times2$ blocks. To see this let us define 
\begin{equation}
M_n:=-
\frac{i}{\pi}nj\left(\begin{matrix}
\frac{a_n}{j} & 0\\
1 & \frac{a_n}{j}
\end{matrix}\right)
\end{equation}
Then, the symplectic form can be cast as an infinite matrix,
\begin{equation}
\omega(T_k^1,T_l^2)=\left(
\begin{matrix}
\ddots&&&&&$\reflectbox{$\ddots$}$\\
&0&0&0&M_{-2}^T&\\
&0&0&M_{-1}^T&0&\\
&0&M_1&0&0&\\
&M_2&0&0&0&\\
$\reflectbox{$\ddots$}$&&&&&\ddots\\
\end{matrix}
\right),
\end{equation}
and the inverse of the symlpectic form is,
\begin{equation}\label{inversesymp}
\omega^{-1}_{kl}:=\omega^{-1}(T_k^1,T_l^2)=\left(
\begin{matrix}
\ddots&&&&&$\reflectbox{$\ddots$}$\\
&0&0&0&(M_{2})^{-1}&\\
&0&0&(M_{1})^{-1}&0&\\
&0&(M_{-1}^T)^{-1}&0&0&\\
&(M_{-2}^T)^{-1}&0&0&0&\\
$\reflectbox{$\ddots$}$&&&&&\ddots\\
\end{matrix}
\right)
\end{equation}
From (\ref{inversesymp}) we can compute the Poisson brackets of $f_n$ and $\alpha_n$,
\begin{eqnarray}\label{poissonbrackets}
\left\{f_n,f_m\right\}&=&0\\
\left\{f_n,\alpha_m\right\}&=&-i\frac{\pi}{na_n} \delta_{n+m}\\
\left\{\alpha_n,\alpha_m\right\}&=&i\frac{\pi j}{na_n^2} \delta_{n+m}
\end{eqnarray}
An important observation is that these brackets imply that the coordinates are of order $1/\sqrt{c_2}$, while $p$ and $j$ are on equal footing as $c_2$ so they are thought as being of order $c_2$. This means that we can consistenly think we are doing a $c_2>>1$ aproximation when we work in a neighborhood of some point in the orbit. On the other hand, we can see the $\alpha$ coordinates as bosons coupled to the $f$ coordinates. In order to decouple the system we could make a change of variables
\begin{equation}\label{coordinates}
s_n=f_n+\frac{a_n}{j}\alpha_n,\qquad t_n=\frac{a_n}{j}\alpha_n,
\end{equation}
and now the Poisson brackets are,
\begin{eqnarray}
\left\{s_n,t_m\right\}&=&0\\
\left\{s_n,s_m\right\}&=&-i\frac{\pi}{nj} \delta_{n+m}\\
\left\{t_n, t_m\right\}&=&i\frac{\pi}{nj} \delta_{n+m}
\end{eqnarray}
Note the resemblance of the $s$-bosons with the ones of Virasoro orbits studied in \cite{Witten}. Nevertheless, we shall not use these coordinates in what follows. 

\subsection{The observables on BMS orbits}

We are interested in computing the explicit form of the classical observables defined over the coadjoint orbits, $\Phi_{u}(\cdot)=\langle\cdot,u\rangle$, which are labeled by an adjoint vector $u$. An element on the orbit of the coadjoint vector $b=(j,0;p,ic_2)$, with $p$ and $j$ constant, is given by the coadjoint action Ad$^*$ of BMS$_3$\footnote{We use the notation $(f,\alpha)$ to express an element of BMS$_3$ group, although the correct way should be $(f,\text{a};\alpha,\text{b})$ where $a$ and $b$ are real numbers and belong to the centrally extended versions of Diff$(S^1)$ and Vect$(S^1)$ respectevely.}
\begin{equation}
\text{Ad}_{(f,\alpha)^{-1}}^*(j,0;p,ic_2)=(\tilde j,0;\tilde{p},ic_2),\qquad f\in \text{Diff}(S^1),\,\alpha\in\text{Vect}(S^1),
\end{equation}
where, having in mind that $c_1=0$,
\begin{eqnarray}
\tilde{j}&=& (f')^2\left[j+\alpha p' +2\alpha' p- \frac{c_2}{24\pi} \alpha'''\right]\circ f\\
\tilde{p}&=& (f')^2p\circ f-\frac{c_2}{24\pi} \left(\frac{f'''}{f'}-\frac{f''^2}{f'^2}\right)
\end{eqnarray}
Let us start with the observable associated with time translations, i.e. the energy $P_0=\Phi_{(0,0;1,0)}=\int \tilde{p} d\phi$, of a generic coadjoint element $(\tilde j,0;\tilde{p},ic_2)$. Writing $f=\theta + \frac{1}{2\pi}\sum_{n\neq 0} f_n e^{-in\theta}$ and $\alpha=\frac{1}{2\pi}\sum_{n\neq 0} \alpha_n e^{-in\theta}$,  we get,
\begin{equation}
P_0=2\pi p+ \frac{1}{\pi}\sum_{n\geq 1}a_nn^2f_n f_{-n}+\mathcal{O}(3)
\end{equation}
Similarly, for the operator associated to rotations $J_0=\Phi_{(1,0;0,0)}=\int \tilde{j} d\theta$ we find,
\begin{equation}
J_0=2\pi j+ j\frac{1}{\pi} \sum_{n\geq 1} n^2 \left(f_n f_{-n} +\frac{a_n}{j}f_n \alpha_{-n}+\frac{a_n}{j} \alpha_nf_{-n}\right)+\mathcal{O}(3)
\end{equation}
The remaining observables can be put in terms of the coordinates also \footnote{We are using the convention of \cite{BarnichOblak1}, where $J_m$ is the observable associated to the vector $(e^{im\phi},0;0,0)$ and analogously  $P_m$ is the observable associated to the vector $(0,0;e^{im\phi},0)$.},
\begin{eqnarray}
J_n&=&-2i\,j n(f_n+\frac{a_n}{j}\alpha_n)\\
P_n&=&-2ina_nf_n
\end{eqnarray}
where $n\neq 0$ and it is clear that they are of order $\sqrt{c_2}$. The set of observables in terms of the coordinates satisfy the \textsf{bms}$_3$ algebra to leading order.

\subsection{The perturbative quantization of BMS and its characters}

In this section we will mimic the approach we followed for computing the characters of Poincar\'{e}, basically using Darboux-like coordinates on the orbit. Recall that a general element in the orbit with little group $S^1\times \mathbb{R}$ can be written as Ad${}_{(f,\alpha)}^*[(j,0;p,c_2)]$, where  
$$f(\theta)=\frac{1}{2\pi}\sum_{n\neq 0}f_n e^{-in\theta},\qquad \alpha=\frac{1}{2\pi}\sum_{n\neq 0}\alpha_n e^{-in\theta},$$ 
so our coordinates in the orbit close to the point $(j,0;p,c_2)\in\,$\textsf{bms}${}_3^*$ are $\left\{f_n,\alpha_m\right\}_{n,m\in \mathbb{Z}-\left\{0\right\}}$. They satisfy $f_n^*=f_{-n}$ and $\alpha_{n}^*=\alpha_{-n}$ and also the Poisson brackets (\ref{poissonbrackets}). Motivated by the procedure for the Poincar\'{e} group as well as the fact that coadjoint orbits of semidirect products $F\ltimes A$ are diffeomorphic to the cotangent bundle of orbits of $A^*$ under the action of $F$, we make the following change of coordinates
\begin{eqnarray}
x_n&:=&\frac{1}{2}(f_n+f_{-n}),\qquad y_n:=\frac{i}{2}(f_n-f_{-n}),\\
p_n&:=&\frac{in}{\pi} \left(\frac{j}{2}(f_n-f_{-n}) + a_n (\alpha_n-\alpha_{-n})\right),\\
q_n&:=&\frac{n}{\pi} \left(-\frac{j}{2}(f_n+f_{-n}) - a_n (\alpha_n+\alpha_{-n})\right),
\end{eqnarray}
where $n>0$. These coordinates, which are real, have canonical Poisson brackets 
\begin{equation}
\left\{x_n,p_m\right\}=\delta_{n-m}, \quad \left\{y_n,q_m\right\}=\delta_{n-m}
\end{equation}
and thus, upon quantization, 
\begin{equation}\label{algebra}
\left[x_n,p_m\right]=i\delta_{n-m}, \quad \left[y_n,q_m\right]=i\delta_{n-m}
\end{equation}
For given $n$ and comparing to the Poincar\'{e} case, $x_n$ plays the part of $x_1$ and $y_n$ plays the part of $x_2$, and similarly their conjugate momenta. 

The observables $P_0$ and $J_0$ read, 
\begin{eqnarray}\label{operators}
P_0&=&2\pi p+ \frac{1}{\pi}\sum_{n\geq 1} a_n n^2 \left(x_n^2+y_n^2\right),\nonumber\\
J_0&=&2\pi j- \sum_{n\geq 1} n\left(x_n q_n -y_n p_n\right). 
\end{eqnarray}
There is no ambiguity when considering these observables as operators on a Hilbert space since $x_n$ and $q_m$ commute for every integer $n,m\geq 1$ and the same happens with $y_n$ and $p_m$. The remaining observables are written as,
\begin{eqnarray}
J_n&=&-ijn(x_n-iy_n)-\pi(p_n-iq_n),\quad n>0\\
P_n&=&-2ina_n(x_n-iy_n),\quad n>0, 
\end{eqnarray}
and $J_{-n}=J_n^*$, $P_{-n}=P_n^*$.

We can represent the operator algebra (\ref{algebra}) on the tensor product of representations of the 2+1 Poincar\'{e} group, as in the previous section. Put differently, the Hilbert space for BMS$_3$ associated with the coadjoint orbit of $(j,0;p,c_2) \in \textsf{bms}_3^*$ will be the product of infinite Hilbert spaces of the two-dimensional particle:
\begin{equation}\label{Hilbert}
\mathcal{H}=\bigotimes^{\infty}_{n=1} L^2(\mathbb{R}^2)_{(n)},
\end{equation} 
where $L^2(\mathbb{R}^2)_{(n)}$ means the Hilbert space associated to the particle of the $n$-th coordinates. The exponential of the operators in (\ref{operators}) is then a tensor product of operators (\ref{rotationandtranslation}), each of them acting on the different $L^2(\mathbb{R}^2)_{(n)}$ Hilbert spaces.  

The representation is then equivalent to the one of infinite non-interacting particles in $2+1$ dimensions and we are considering the non-relativistic limit where the energy is quadratic. With this we mean that if again we choose a basis of plane waves $\psi_{\left\{\vec k\right\}}:=\bigotimes_{n\geq 1}\psi_{\vec{k}_n}$, the energy is quadratic in each $\vec{k}_n$. In addition, as mentioned earlier, the approximation can be seen as a $c_2\rightarrow\infty$ limit (or $p>>1$). Thus, $c_2$ plays the part of speed of light.     

It is worth mentioning the relation to the induced representations used in \cite{Oblak}. If we relate an array of vectors $\left\{\vec{k}_n\right\}_{n\geq 1}$ (a choice of plane waves) with a point in the orbit Ad$^*_{\text{Diff}(S^1)}(p t^{0*})=\text{Diff}(S^1)/S^1$, which is a plane wave in \cite{Oblak}, then the relation to induced representations of \cite{Oblak} starts to become clear. Moreover, the rotations and pure translations of the induced representations act in the same way as in our coadjoint orbit representations, by rotating the plane wave or adding a complex phase, respectively \footnote{It remains to see if any exponential of the algebra (perturbative) representation matches the induced representations, although this is likely the case.}.

We have now arrived to a position where we can calculate the characters of these representations: first note that the coordinate-independent terms in the operators in (\ref{operators}) will give again a phase, and thus what is left is an infinite product of characters of Poincar\'{e} (not including the phase just mentioned). However, for each Poincar\'{e} character (\ref{poincarecharacter}) one needs to take into account that its corresponding rotation now comes weighted by the integer $n$ in (\ref{operators}),
\begin{equation}
\chi_{(j,p,c_2)}\left(e^{i\theta J_0+i\beta P_0}\right)=e^{i2\pi\theta j+i2\pi\beta p}\prod_{n\geq 1}\frac{1}{|1-q^n|^2},\qquad q=e^{i\theta}.
\end{equation}
This matches the character computed in \cite{Oblak}, having in mind a slight difference of normalization in the operators: in that reference for example $P_0$ comes with a $(2\pi)^{-1}$ normalization so its classical value is just $p$. 

It remains to consider the case where $p=-\frac{c_2}{48\pi}$, where the orbit is now the cotangent space of the Virasoro Diff$(S^1)/\text{PSL}(2,\mathbb{R})$ orbit. There is no substantial difference, actually: one only has to take into account that the $\alpha_{\pm 1}$ and $f_{\pm 1}$ modes are now spurious, in the sense that they belong to the isotropy group. Thus, sums and products start from $n=2$ and everything goes straight forward. The character is then
\begin{equation}
\chi_{(j,p,c_2)}\left(e^{i\theta J_0+i\beta P_0}\right)=e^{i2\pi\theta j-i\beta \frac{c_2}{24}}\prod_{n\geq 2}\frac{1}{|1-q^n|^2},\qquad q=e^{i\theta},
\end{equation}
in accordance with \cite{Oblak,Barnichetal}.
 
\section{Conclusions}

The main result of this work is the computation of particular characters of the BMS $_3$ group, following the same lines as those of Witten for the Virasoro group \cite{Witten}. The characters are associated to `coadjoint orbit representations'. We showed how to compute characters for the massive coadjoint orbits of BMS$_3$ group as well as the vacuum orbit. As a byproduct, the (approximate) representations found here were related to the induced representations of \cite{BarnichOblak1,Oblak}, as discussed after equation (\ref{Hilbert}). This is an explicit example of a  relation between induced representations and (perturbative) `coadjoint orbits representations' for an infinite-dimensional Lie group. The characters, as expected, coincide with the ones computed in \cite{Oblak}, were the induced representation was used. 

The reason behind the coincidence between the characters coming from a perturbative approach with the ones of the  induced representations is elusive to us. The fact that the   `coadjoint representation' coincides with the induced representation is valid at the level of the approximation $c_2>>1$, as far as we could show here. However, it is likely to be the case that the coincidence of the characters is not only exact, but also that there is a solid reason coming from general results on representations of infinite-dimensional Lie groups.   

One step in the direction of the last claim is to study the possibility of a generalization of geometric quantization applied to infinite-dimensional groups, as well as its relation with index-theorems. If one could show, without resorting to perturbative methods, that the functional dependence of the character with the central charge (and the other labels $p,j$) is insensitive to the central charge value, then it would be clear why the perturbative approach gives the correct result for all values of the central charge.

As an example of this line of thought, consider computing again the characters of BMS$_3$ but without finding explicitly a representation of the group, only resting on the Lefschetz formula \cite{AtiyahBott}. The explicit original formula relates a geometrical index with a topological index, both associated to a particular map (which in our case is the unitary representation of a group element). The topological index is the Lefschetz number, which reduces to the character when the higher cohomolgy groups vanish (the remaining one would be the space of global sections on some bundle). In order to apply Lefschetz formula some mathematical structures need to be taken into account, but for Virasoro and BMS$_3$ groups these structures are have not been completely studied yet, as far as we know (we will discuss this somewhere else \cite{FarinatiGarbarzLeston}). Nevertheless, this formula was used in \cite{Witten} for the Virasoro group in an intuitive way, and here we follow those lines with suitable slight modifications in order to recover the characters of BMS$_3$ group. First we write the operators $J_0$ and $P_0$ as vectors on the orbit,  
\begin{equation}
P_0=\sum_{n\geq 1} \frac{2a_n n^2}{\pi} \left(x_n \frac{\partial}{\partial p_n}+y_n\frac{\partial}{\partial q_n}\right),\qquad J_0=\sum_{n\geq 1} n\left(x_n\frac{\partial}{\partial y_n} - y_n\frac{\partial}{\partial x_n}+p_n\frac{\partial}{\partial q_n} - q_n\frac{\partial}{\partial p_n}\right)
\end{equation}
Second, we compute finite transformations generated by them, $\exp(i\theta J_0+i\beta P_0)$: for a given polarization, say $\left\{x_n,y_n\right\}$,
\begin{equation}
\left(
\begin{matrix}
x'_n \\
y'_n \\
\end{matrix}
\right)=\left(
\begin{matrix}
\cos\theta & \sin\theta\\
-\sin\theta & \cos\theta\\
\end{matrix}
\right) \left(
\begin{matrix}
x_n\\
y_n\\
\end{matrix}
\right)
\end{equation}
Had we chosen the polarization $\left\{p_n,q_n\right\}$ nothing would have changed. The Jacobian of the transformation, $\mathcal{J}$, is the $2\times2$ rotation matrix above. The last step in the heuristic Lefschetz formula says we should compute det$(1-\mathcal{J})$, 
\begin{equation}
\text{det}(1-\mathcal{J})=\prod_{n=1}^{\infty} |1-e^{i\theta n}|^2  
\end{equation}
and that the character is just the operators evaluated at the classical values times the inverse of $\text{det}(1-\mathcal{J})$,
\begin{equation}
\text{Tr}(e^{i\beta P_0+i\theta J_0})=e^{i\beta p} q^j \prod_{n=1}^{\infty} \frac{1}{(1-q^n)^2},\quad q=e^{i\theta}.
\end{equation}

\section*{Acknowledgments}

We are grateful to G. Giribet and H. Gonzalez for enlightening discussions, and specially to B. Oblak for providing many clarifications about his work on characters of BMS$_3$. This work was mainly supported by CONICET and the University of Buenos Aires, as well as partially supported by grants PIP and PICT from CONICET and ANPCyT, and a MINCyT-FNRS 2013 bilateral agreement.


\begin{thebibliography}{999}

\bibitem{GoodmanWallach2} R. Goodman, N. R. Wallach, \textit{Projective Unitary Positive-Energy Representations of Diff$(S^1)$}, Journal of Funct. Analysis \textbf{63}  (1985) 299.

\bibitem{GarbarzLeston} A. Garbarz, M. Leston, \textit{Classification of Boundary Gravitons in AdS$_3$ Gravity}, JHEP \textbf{1405} (2014) 141; arXiv:1403.3367.
 

\bibitem{BarnichOblak} G. Barnich, B. Oblak, \textit{Holographic positive energy theorems in three-dimensional gravity}, Class. Quant. Grav. \textbf{31} (2014) 152001; arXiv:1403.3835.
 


\bibitem{Witten} E. Witten, \textit{Coadjoint Orbits of the Virasoro Group}, Commun. Math. Phys. \textbf{114}  (1988) 1. 

\bibitem{Balog} J. Balog, L. Feher, L. Palla, \textit{Coadjoint orbits of the Virasoro algebra and the global
Liouville equation}, Int. J. Mod. Phys. A\textbf{13} 315 (1998); arXiv:hep-th/9703045.


\bibitem{BarnichOblak1} G. Barnich, B. Oblak,\textit{Notes on the BMS group in three dimensions: I. Induced representations}, JHEP \textbf{1406} (2014) 129; arXiv:1403.5803.
 
\bibitem{BarnichOblak2} G. Barnich, B. Oblack, \textit{Notes on the BMS group in three dimensions: II. Coadjoint representation}, JHEP \textbf{1503} (2015) 033; arXiv:1502.00010.


\bibitem{BTZ} M. Ba\~nados, C. Teitelboim, J. Zanelli, \textit{The Black hole in three-dimensional space-time}, Phys. Rev. Lett. \textbf{69}  (1992) 1849; arXiv:hep-th/9204099.\\
M. Ba\~nados, M. Henneaux, C. Teitelboim, J. Zanelli, \textit{Geometry of the (2+1) black hole}, Phys. Rev. D \textbf{48}  (1993) 1506; arXiv:gr-qc/9302012. 

\bibitem{NeebSalmasian} K-H. Neeb, H. Salmasian, \textit{Classification of positive energy representations of the Virasoro group}, arXiv:1402.6572.

\bibitem{AtiyahBott} M. F. Atiyah, R. Bott, \textit{A Lefschetz fixed-point formula for elliptic differential operators}, Bull. Amer. Math. Soc. \textbf{72} 2 (1966) 245.

\bibitem{conical} J. Raeymaekers, \textit{Quantization of conical spaces in 3D gravity}, JHEP \textbf{1503} (2015) 060 ; arXiv:1412.0278. 

\bibitem{JoosSchrader} H. Joos, R. Schrader, \textit{On the primitive characters of the Poincar\'{e} group}, Comm. Math. Phys. \textbf{7} 1 (1968) 21.

\bibitem{FuchsRenouard} G. Fuchs, P. Renouard, \textit{Characters of Poincar\'{e} group}, J. Math. Phys. \textit{11} (1970) 2617.

\bibitem{Oblak} B. Oblak, \textit{Characters of the BMS Group in Three Dimensions}; arXiv:1502.03108.

\bibitem{Barnichetal} G. Barnich, H. Gonzalez, A. Maloney, B. Oblak, \textit{One-loop partition function of three-dimensional flat gravity }, JHEP \textbf{1504} (2015) 178; arXiv:1502.06185. 

\bibitem{Wigner} E. P. Wigner, \textit{On unitary representations of the inhomogeneous Lorentz group}, Annals Math. \textbf{40} (1939) 149.

\bibitem{FarinatiGarbarzLeston} M. Farinati, A. Garbarz, G. Giribet, M. Leston, in preparation.

\end{thebibliography}
\end{document}